\documentclass[prl,aps,twocolumn, floatfix, superscriptaddress,showpacs]{revtex4}
\usepackage{amsmath,bm,graphicx}
\usepackage{amssymb}
\usepackage{amsfonts}
\begin{document}

\title{Entropic electrokinetics: recirculation, particle separation and negative mobility}
 
\author{Paolo Malgaretti}
\email[Corresponding Author : ]{paolomalgaretti@ffn.ub.es }
\affiliation{Department de Fisica Fonamental, Universitat de Barcelona, Carrer Mart\'{\i} i Franqu\'es, 08028-Barcelona, Spain}
\author{Ignacio Pagonabarraga}
\affiliation{Department de Fisica Fonamental, Universitat de Barcelona, Carrer Mart\'{\i} i Franqu\'es, 08028-Barcelona, Spain}
\author{J. Miguel Rubi}
\affiliation{Department de Fisica Fonamental, Universitat de Barcelona, Carrer Mart\'{\i} i Franqu\'es, 08028-Barcelona, Spain}
\affiliation{Department of Chemistry, Imperial College London, SW7 2AZ, London, UK}
\date{\today}

\begin{abstract}
We show that when particles are suspended in an electrolyte confined between corrugated charged surfaces, electrokinetic flows lead to a new set of phenomena such as particle separation, mixing for low-Reynolds micro- and nano-metric devices and negative mobility. Our analysis shows that such phenomena arise, for incompressible fluids, due to the interplay  between the electrostatic double layer and the corrugated geometrical confinement and that they are magnified when the width of the channel is comparable to the Debye length. Our characterization allows us to understand the physical origin of such phenomena therefore shading light on their possible relevance in a wide variety of  situations, ranging from nano- and micro-fluidic devices to biological systems. 
\end{abstract}



\maketitle

The recent development of nano- and micro-fluidic devices~\cite{Lyderic} as well as cellular regulation mechanisms and cellular signaling~\cite{Albers} rely on the transport of ions across channels or pores whose sections range from the nanometric to the micrometric scale~\cite{Karnik2007,Park2006,Lyderic_nature}. The transport across such conduits has been characterized, even for varying-section channels~\cite{hanggi,ghosal,Dorfman2008,Adler,Reguera2006, Umberto_jcp}, assuming that the channel width, $h(x)$,  is  large compared to the Debye length, $\kappa^{-1}$, over which the  electrolyte charge distributes in the neighborhood of the charged channel wall ($\kappa h(x)\gg 1$), or in the absence of electrolytes~\cite{Hanggi_PRL}.
Nowadays, the continuous process of device miniaturization and the widening of the range of achievable salt concentrations require an understanding of the behavior of such systems when the relevant length scales compete with each other~\cite{Rotenberg_review}. Such regimes are already exploitable in different micro- and nano-fluidic experiments~\cite{Karnik2007,Park2006} and can be relevant in a variety of biological systems~\cite{Lemaire2005,Strook2008}. 
\begin{figure}
\includegraphics[scale=0.235]{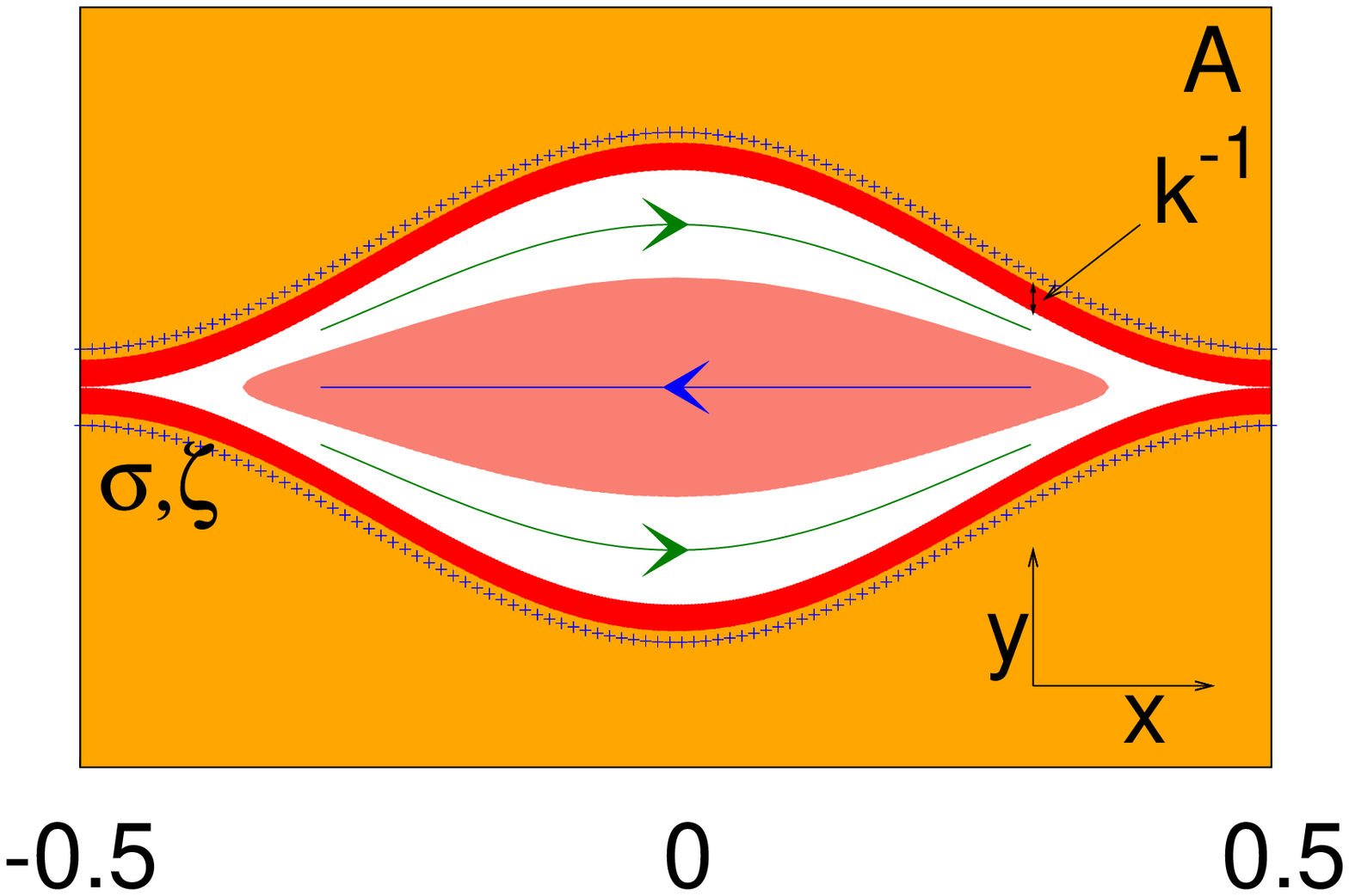}\includegraphics[scale=0.235]{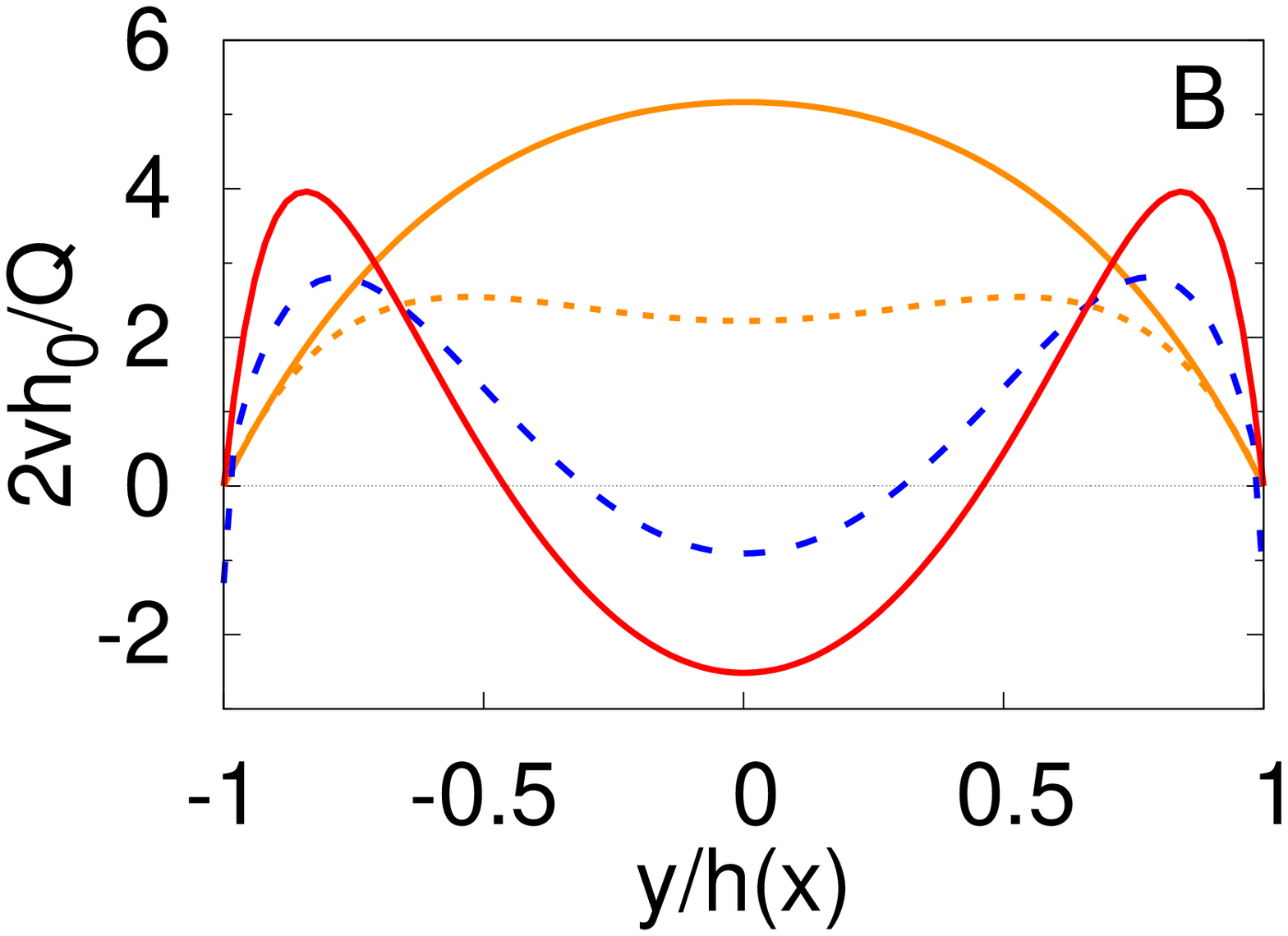}
\includegraphics[scale=0.235]{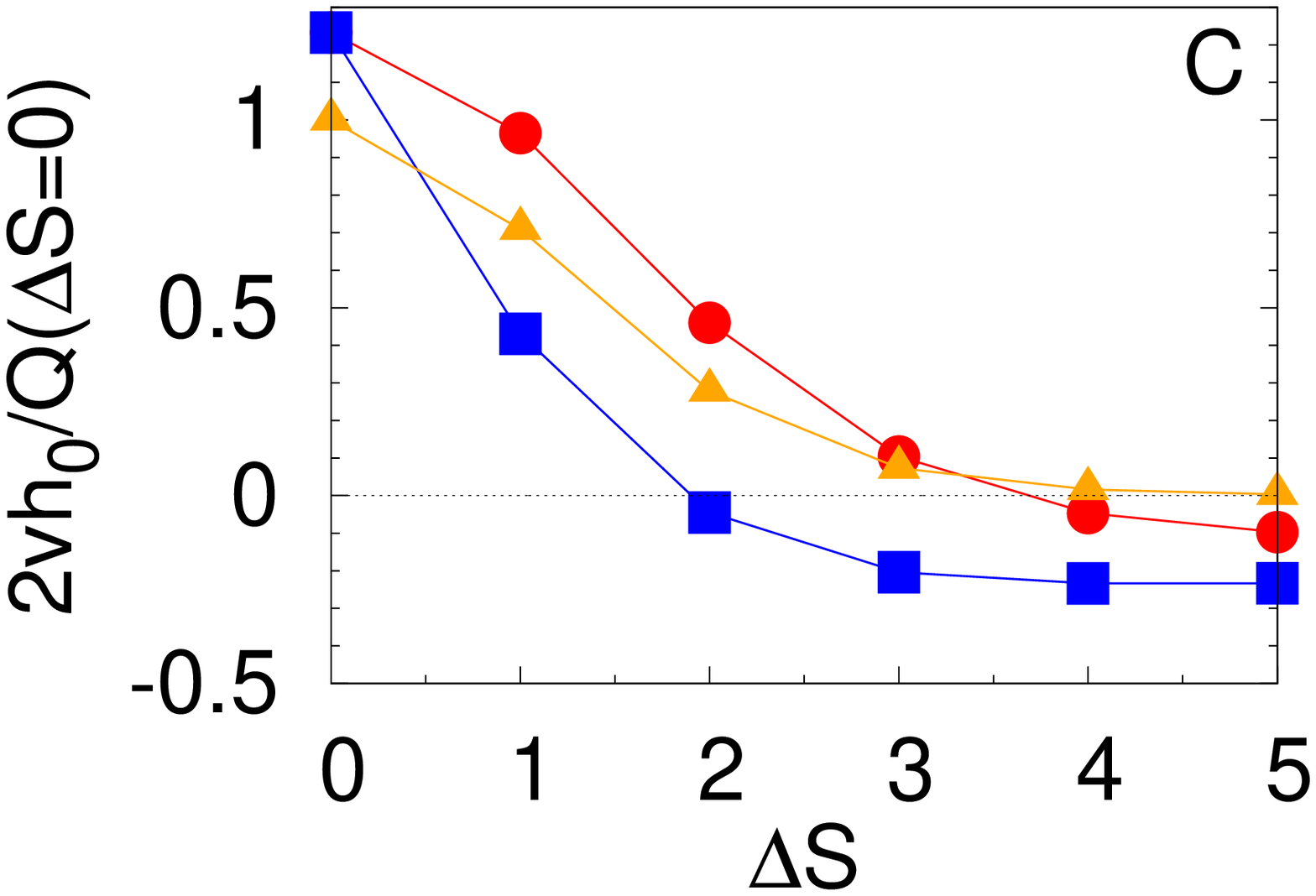}\includegraphics[scale=0.235,bb = 50 75 590 453]{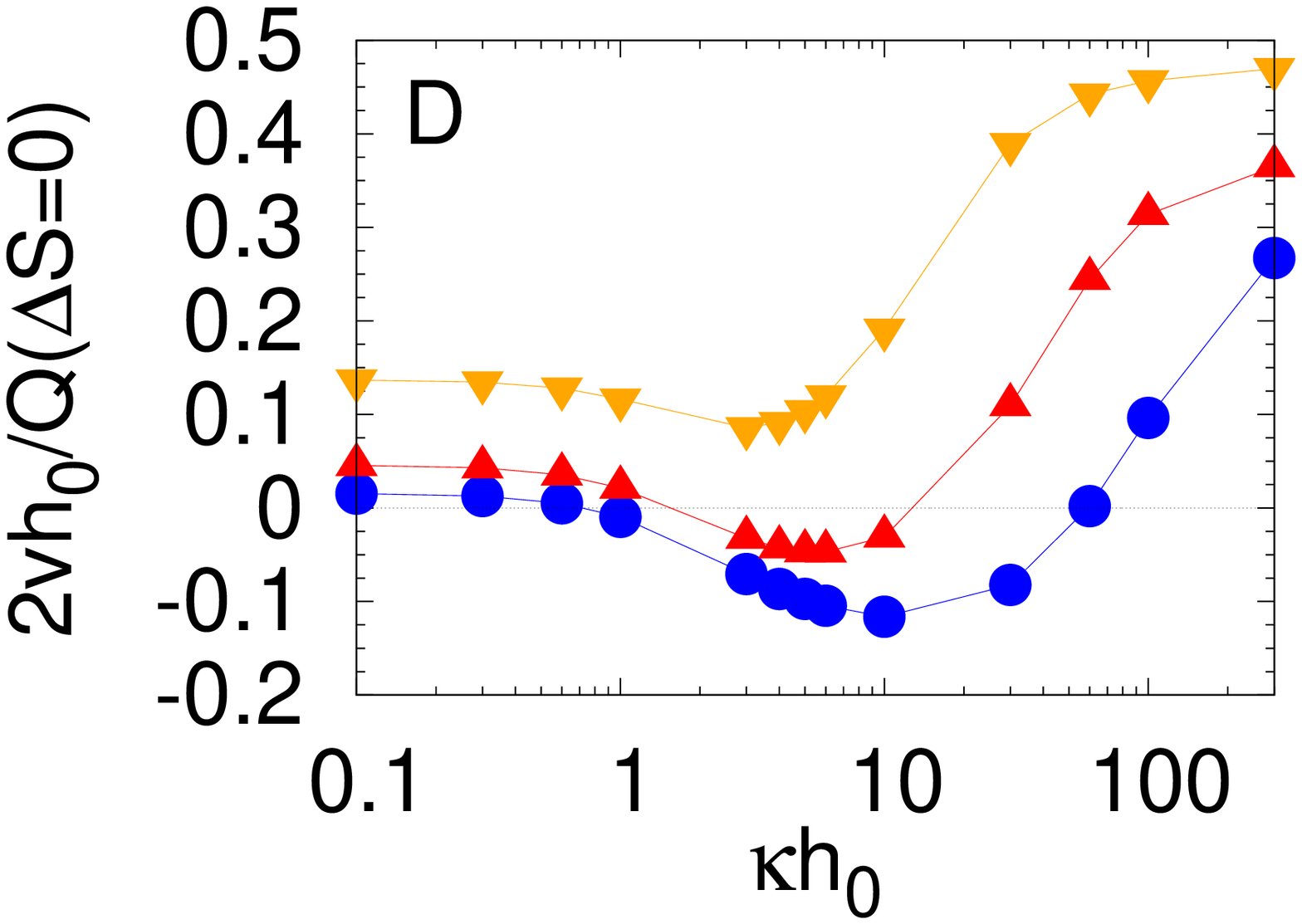}
\caption{Geometry and characteristic flows of  entropic electro-osmotic transport. A: Schematic view of the channel; B:   Velocity profiles for an electrolyte electro-osmotically driven inside a varying-section insulating channel of section $h(x)$ (as described in the text), at three positions along the channel, $x/L=0,0.25,0.35,0.4$ (red solid, blue dashed, orange dotted and red thick, respectively) for $\kappa h_0=5$, $\beta ze \Delta V=0.01$, $\Delta S=3$. C: Mass flow, $Q/(2h_0)$ (triangles), average velocity along the channel longitudinal symmetry axis (circles) and velocity at $x/L=1/2,\, y=0$ (squares), normalized by the corresponding mass flow along a uniform section channel, $Q(\Delta S=0$),  as a function of the corrugation for $\kappa h_0=5$, $\beta ze \Delta V=0.01$. D: average fluid velocity along the longitudinal axis, normalized by the corresponding mass flow along a uniform section channel, $Q(\Delta S=0$), as a function of $\kappa$ for $\beta ze \Delta V=0.01$ and $\Delta S=3$ (orange down 
triangles), $\Delta S=4$ (red up triangles), $\Delta S=5$ (blue circles).}
\label{fig:1}
\end{figure}

In this Letter we will show that, precisely in this regime, i.e. when the Debye length and the channel aperture are comparable in size, $\kappa h(x)\sim 1$, an electrolyte embedded in a corrugated channel develops new transport regimes that can be exploited to separate suspended particles, control electric and mass currents and eventually induce negative mobility.
When $\kappa h(x)\sim 1$, the electrolyte response to external forcing, such as electrostatic fields, is very sensitive to the channel  shape and it develops a recirculating region in which the electrolyte flows on the opposite direction as compared to the average volume flow, as shown in Fig.~\ref{fig:1}.A. Such a phenomenon, typical for incompressible fluids, is due to the interplay between the electrostatic double layer and the varying geometrical confinement and its magnitude is significantly amplified when $\kappa h(x)\sim 1$. We coin this regime entropic electrokinetics since the phenomena we identify can only arise due to the spatially varying constriction induced by the geometrical confinement. This variation affects the local spatial distribution of ions, essentially controlled by the interplay between the wall charge and the ion entropy. We will show that  the entropic variations in the charge density induced by the geometric constraints of the corrugated walls have a significant influence in the  
electrokinetics  for narrow channels.
Moreover, the dynamics of suspended particles  is strongly affected by the local recirculation of the electrolyte: phenomena such as current inversion particle separation and negative mobility can be attained  appropriately  tuning the channel shape. 
Therefore, entropic electrokinetics can be exploited in diverse situations such as low Reynolds fluid mixers~\cite{Stroock2010}, electrokinetic batteries~\cite{Yang2003}, microfluidic circuitry~\cite{Lyderic_PRL} and salinity-based energy harvesting devices~\cite{Brogioli2009,Roji2011} as well as biologically relevant systems such as transpiration in plants~\cite{Strook2008} or cortical bone fluid flows~\cite{Lemaire2005}.

In order to characterize this electrokinetic transport regime, we will consider a symmetric,  $z-z$ electrolyte 
 solution, in contact with a reservoir  of ionic strength $\rho_0 z^2$, and  filling  a varying-section channel of  length $L$ and half-aperture $h(x)=h_0-h_{1}\cos\left(2\pi x/L\right)$ and whose walls, flat along the $z$ direction, have either a constant surface charge $\sigma$ or a constant electrostatic potential $\zeta$. The geometric impact of the channel corrugation on the electrokinetics of the liquid can be quantified in terms of the entropic barrier $\Delta S=\ln \frac{h_0+h_1}{h_0-h_1}$~\cite{paolo_jcp_2013}. In this highly confined geometry  we consider that the channel aperture varies smoothly, $\partial_x h(x)\ll1$.  This regime, where   $\partial_x^2\ll\partial_y^2$ allows us to take advantage of the  lubrication approximation~\cite{Park2006} and reduce  both the  Poisson-Boltzmann equation governing the electrostatic field and the Stokes equation governing the low-Reynolds number fluid flow to $1D$ differential equations, as described in the Supplementary Material (SM).  In this 
regime we can disregard the dependence of the pressure on 
the transverse  location,  $P(x,y)=P(x)$, and  assume that co- and counter-ions attain their equilibrium profile along the transverse direction, $y$, leading to a charge density
\begin{equation}
 q(x,y)=ze\rho_{+}(x)e^{-\beta ze \phi(x,y)}-ze\rho_{-}(x)e^{\beta ze \phi(x,y)}
 \label{charge-dens}
\end{equation}
where  $\beta^{-1}=k_BT$ corresponds to the inverse thermal energy, $e$ stands for the elementary charge, $z$ for the valence of the electrolyte and $\phi(x,y)$ is the electrostatic potential.  The varying channel aperture  induces a variation of the  overall ionic density along the  channel, quantified by the amplitudes, $\rho_{\pm}(x)$. Within the linearized Debye-H\"uckel  regime, the electrostatic potential in the channel reads
\begin{equation}
\phi(x,y)=\frac{\Lambda}{\kappa^2} \cosh\left(\kappa y\right)+\frac{ze}{\epsilon \kappa^{2}}\left[\rho_{+}(x)-\rho_{-}(x)\right]
\label{eq:el_pot}
\end{equation}
where $\kappa=\sqrt{4 \pi \ell_B z^2(\rho_+(x)+\rho_-(x))}$ is the local  inverse Debye length for an electrolyte of valence $z$, $\ell_B=\beta e^2 /4\pi\epsilon$ stands for the Bjerrum length for an electrolyte  with dielectric constant $\epsilon$ and $\Lambda$ is determined by the boundary condition on the channel walls~\footnote{For insulating channel walls with surface charge $\sigma$,  local electroneutrality is always preserved in the  lubrication regime. Although we disregard  variation of  the electric field  at the  confining wall   due to the channel corrugation (see e.g. ~\cite{Eisenberg}), its influence is of higher order than the results we will analyze.}.

The  electrolyte dynamics is quite sensitive to the boundary conditions on the channel walls (either conducting or insulating) as well as to the nature of the external forcing (pressure or electric field; a detailed discussion can be found in the SM). We will focus on the electroosmosis in a channel with insulating corrugated walls, when the induced dependence of the local electrostatic field along the channel is subdominant. As described in the SM, imposing  constant mass, $J_\rho$, charge, $J_q$, and solvent fluxes along the channel provides the steady state charge density$\rho_{\pm}$, pressure, $P(x)$, and  velocity , $v(x)$, profiles (cf. Eqs.(2-6) in SM). The variable   channel aperture  introduces geometry-dependent constrictions and induces an inhomogeneous pressure gradient that controls the electrokinetic response. In equilibrium, the deviations of the ionic density from the magnitude corresponding to a uniform channel is negligible, $\rho_{\pm}^{(eq)}=\rho_0/2$~\footnote{These results shed light 
on the corrugation-induced correction at equilibrium, stating that up to second order in the surface charge there is no relevant correction to the homogeneous profile $\rho_{\pm}^{(eq)}=\rho_0/2$}

In this regime, due to the channel  smooth spatial corrugation, $\partial_x h(x)\ll1$, the electrolyte  response is  dominated by the local pressure drop, while the contribution from the corrugation-induced corrections to the electrostatic potential is negligible.  As a result, the transport and dynamics of electro- and pressure-driven  electrolytes display qualitative and quantitative differences. On the contrary, for strongly corrugated channels, $\partial_x h(x)\gg1$ we expect the corrections in the ion density to overwhelm the local geometrically-induced pressure gradients leading to a complementary scenario,  beyond the perturbative  approach described. Accordingly, we can identify two corrugation-induced electrokinetic regimes: when $\partial_xh(x)\ll1$  ion densities are well captured by Eq.~\ref{charge-dens} i.e. the dynamics of ions density is well captured by the concept of entropic barriers~\cite{zwanzig,Reguera2001,paolo_jcp_2013}. For  $\partial_x h(x)\gg1$ the factorization in Eq.~\ref{charge-
dens} brakes down and the transverse ion distribution gets more involved. 
\begin{figure*}
\includegraphics[scale=0.32]{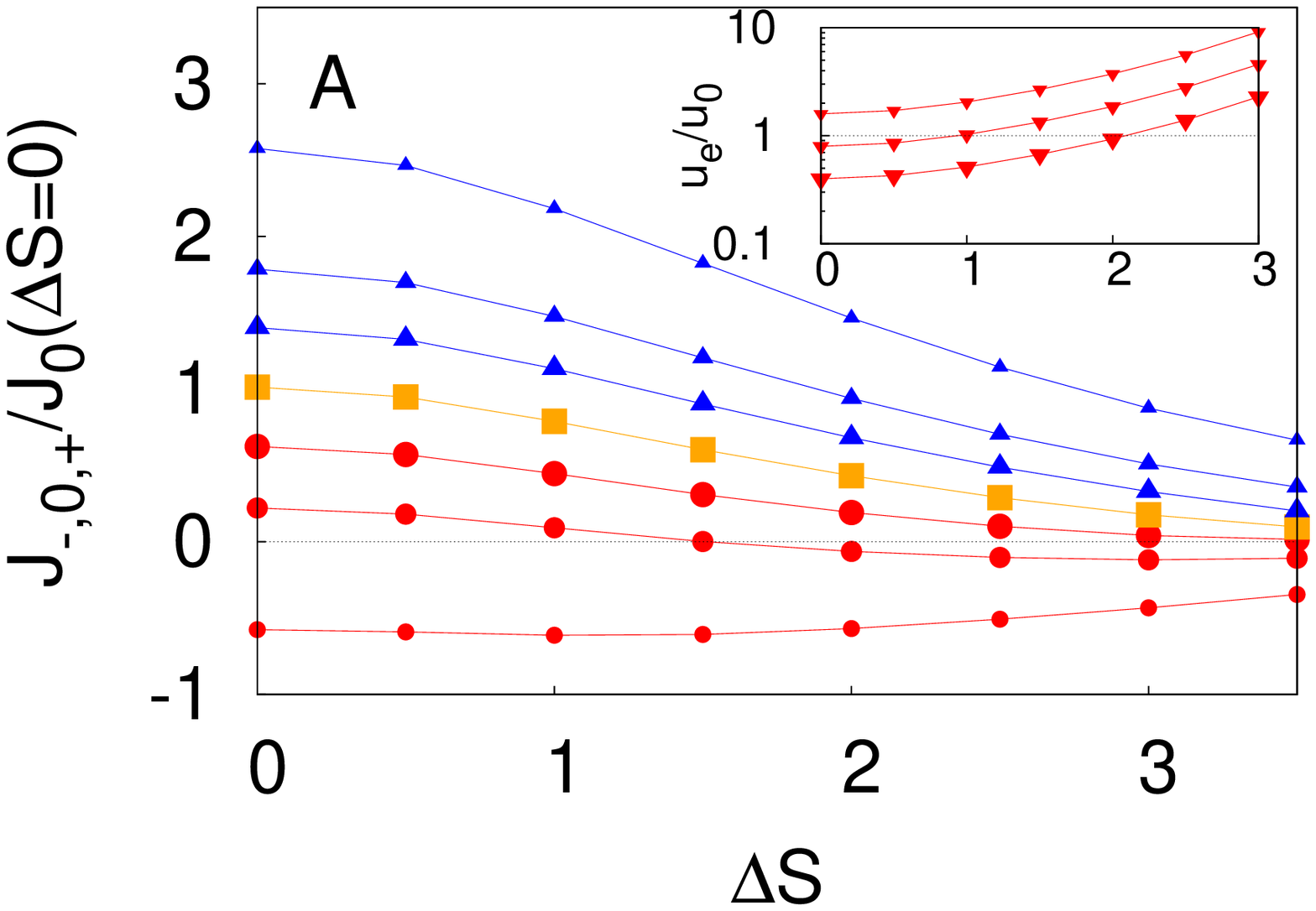}\includegraphics[scale=0.32]{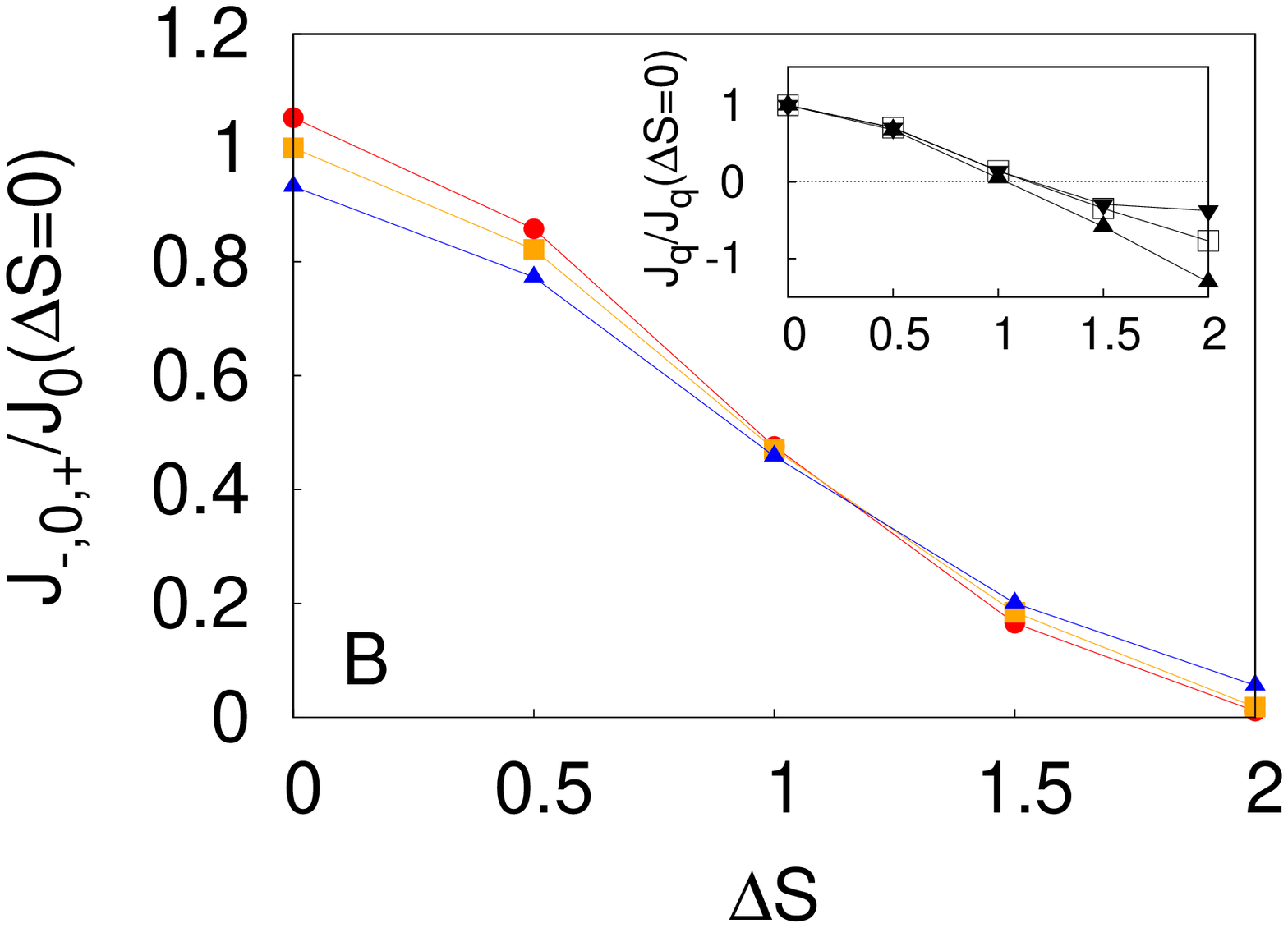}
\includegraphics[scale=0.32]{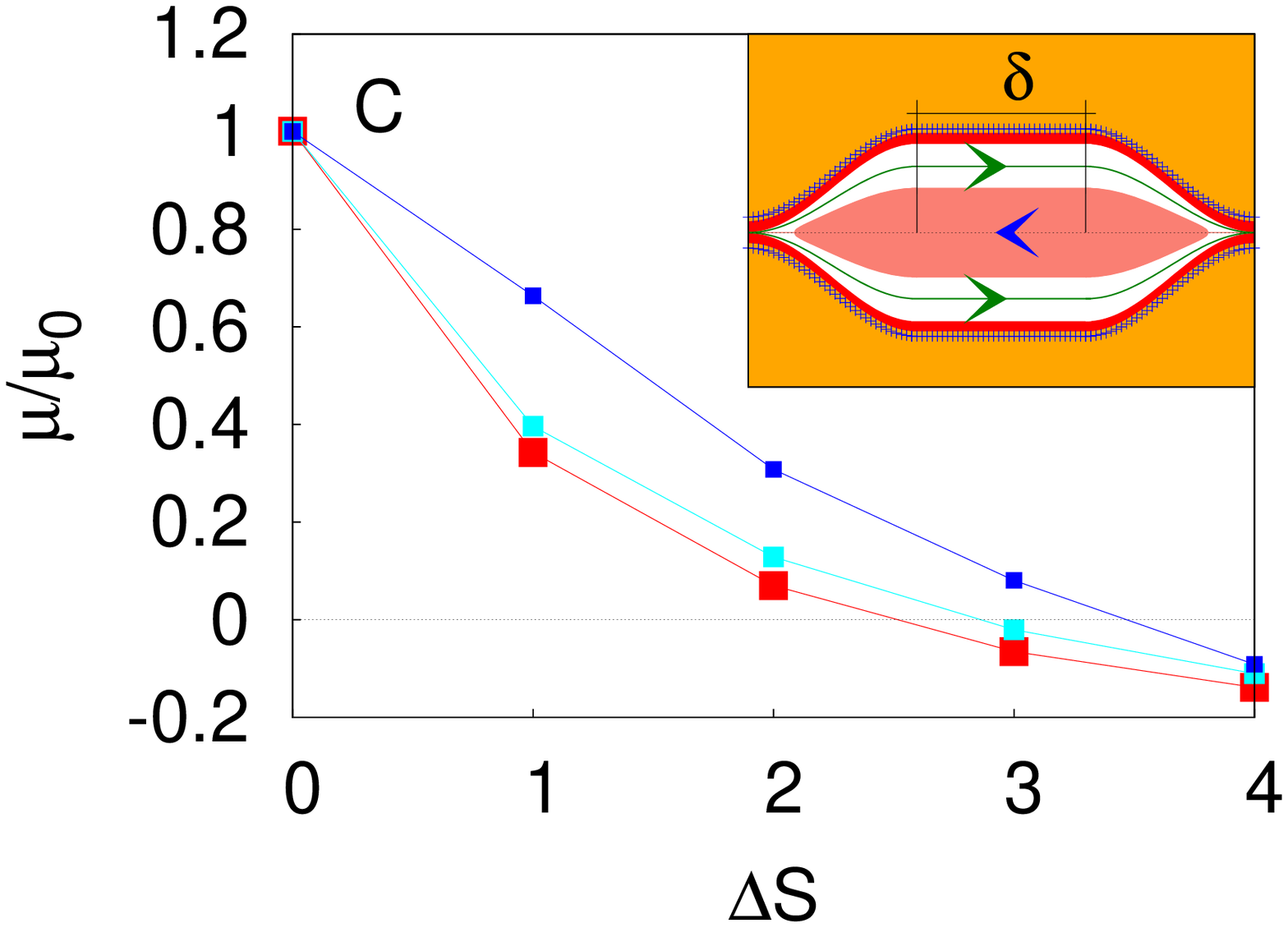}
\caption{Tracer flux in a varying-section channel as a function of the entropy barrier, $\Delta S$. A: positive (red dots), neutral (orange squares), or negative (blue triangles) tracer flux due to external electric field $\beta ze \Delta V=0.01$ characterized by $\kappa h_0=20$ for different tracer radii: $R/L=0.001,0.002,0.004$ where bigger points stand for bigger radii. Inset: ratio between the external electrostatic and the convective force for the same data as in the main figure. B: positive (red dots), neutral (orange squares), or negative (blue triangles) tracer flux due to external  pressure drop $\beta \Delta P h^2_0=0.001$ for tracer radius $R/L=0.004$. Inset: electric current, $J_q$, normalized by its value at $\Delta S=0$ with $R/L=0.001,0.002,0.004$ for upper triangles, open squares, and lower triangles,  respectively. C: Mobility, $\mu$, of neutral tracers confined to the  center of a corrugated channel, normalized by the mobility for a flat channel, $\mu_0$,  as a function of $\Delta S$  for 
$R/L=10^{-4}$ and $\delta=1,2,3$ where bigger points stand for larger values of $\delta$.}
\label{fig:tracers}
\end{figure*}

Fig.~\ref{fig:1}.B displays the velocity profile of the electrolyte to leading order in  the applied voltage drop, $\Delta V$, (corresponding to an applied uniform electric field, $E_x=-\Delta V/L$) along the channel  when the  variable channel section   does not distort significantly the equilibrium electrostatic potential, $\phi_0$. Interestingly, in a corrugated channel characterized by a large entropic barrier, the electro-osmotic velocity in the central region of the channel  flows against the mean fluid flow, $Q$. This velocity inversion appears due to the mismatch between the electroosmotic fluid profile and the  confinement-induced Poiseuille flow, and grows with the geometric mismatch, or entropy barrier, $\Delta S$, as displayed in Fig.~\ref{fig:1}.C.
In particular, fig.\ref{fig:1}.D shows that velocity reversal is magnified and reaches a significant magnitude when $\kappa^{-1}$ is comparable to the typical channel width, highlighting that entropic electrokinetic transport emerges from a competition between the reorganization in the charge profiles due to the geometrical confinement and its impact in the electroosmotic and  pressure-driven flows, characteristic of the entropic electrokinetic regime.  Although  we have focused in the linearized electrostatic regime to emphasize the origin of  geometrically controlled electrolyte transport,   the geometric restrictions induced by  corrugated channels  are robust and will  affect electrokinetics   also in strongly coupled~\footnote{Since the confined electrokinetic regime relies on the flow inversion stemming from incompressibility, we expect it to be robust  for strongly coupled electrolytes} and strongly forced electrolytes~\footnote{The corrections due to the imposed electric field  arise through the 
deformations it induces  in the ionic distribution. The magnitude of the electric field parallel to the channel  is of order $F_{ext} \simeq ze E\partial_xh$ while the electric field due to the charged wall has a typical magnitude $F_{Dby} \simeq ze\phi_0\kappa$. Therefore,  the distortion of the charge profile along the channel is negligible when $E\ll \kappa\phi_0/\partial_x h$, which is easily achieved in narrow or slowly varying channels,  $\partial_x h\ll 1$.}.  Entropic electrokinetics, relevant to promote  particle mixing  at the micro and nanoscale as well as for controlling electric currents, is  more prominent  for insulating  walls~\footnote{For conducting  walls velocity reversal develops for strong corrugations (data not shown)} and vanishes for pressure driven flows~\footnote{For 
pressure driven flows, the matching between the driven and the induced flux affects more mildly the overall flow not giving rise to flow inversion in the center of the channel.}.

The  electrokinetic flows induced by corrugated charged channels have a strong effect on the dynamics of suspended tracers. Tracers, quantified in terms of their local concentration, $C(x,y,t)$, will diffuse (with diffusion coefficient $D_{tr}$), will be advected according to the fluid  velocity, $v_x(x,y)$, and will be forced by the local electric field depending  to the tracer charge.  Following the Fick-Jacobs approximation~\cite{zwanzig,Reguera2001,Kalinay2008} that has been shown to properly capture the dynamics of suspended tracers in corrugated channels~\cite{Reguera2001} when $\partial_x h(x)\ll 1$, we can assume that   tracers retain their equilibrium distribution along the transverse direction, and factorize $C(x,y,t)=p(x,t)g(y|x)$, with 
\begin{equation}
g(y|x)=\frac{e^{-\beta U(x,y)}}{e^{-\beta A(x)}},\, e^{-\beta A(x)}=\int_{-\infty}^{\infty}e^{-\beta U(x,y)}dy
\label{fick-jacob}
\end{equation}
in terms of the effective potential
$$U(x,y)=\left\{
\begin{array}{cc}
Ze (\phi(x,y)-x\tilde{E}_{x,0}) & y\in[-h(x),h(x)]\\ \infty & \mbox{otherwise}
\end{array}
\right.$$
being $Z$ the tracer valency and $p(x,t)$ the probability of finding a tracer at position $x$ at time $t$. For dilute tracers, we can disregard tracer-tracer interactions and  tracer-induced corrections in the electrostatic fields and  pressure gradient. Accordingly, tracer dynamics can be expressed in terms of an  effective, $1D$  convection-diffusion equation
\begin{equation}
\frac{\partial}{\partial t} p=\frac{\partial}{\partial x}\left[\left(\frac{\partial}{\partial x} D_{tr}\beta A(x)-\left\langle v_{x}\right\rangle _{y}\right)p+D_{tr}\frac{\partial}{\partial x} p\right]
\label{eq:final_FP}
\end{equation}
where $\left\langle v_{x}\right\rangle _{y}=\frac{1}{2h(x)}\int_{-h(x)}^{h(x)}v(x,y)e^{-\beta U(x,y)}dy$ stands for the average velocity over the channel section. Eq.~\ref{eq:final_FP} already shows that the tracer dynamics is governed by the average velocity field $\left\langle v_{x}\right\rangle _{y}$ and by the entropic barrier, $A(x)$, that according to Eq.~\ref{fick-jacob} has the shape of the equilibrium tracer free energy.  

Fig.~\ref{fig:tracers} shows that tracer velocities, generically, decrease due to the entropic barriers induced by  channel corrugation  both for electrically and pressure-driven  driven electrolytes for insulating walls~\footnote{Analogous results are obtained for conducting walls, although of smaller magnitude (everything else being equal)}.
In electroosmosis in a flat channel, $\Delta S=0$, negatively charged tracers (closer to the positively charged surface) benefit from the forcing imposed by the external field and move faster than both positive and neutral ones~\cite{Rotenberg_epl}.  Fig.~\ref{fig:tracers}.A shows that channel corrugation generically  decreases asymmetrically the electroosmotically-induced tracer velocities. Increasing $\Delta S$ the fluid velocity  develops a region of counterflow (as seen in  Fig.\ref{fig:1}), leading to a reduction of the convective contribution to the tracer motion with respect to the electrostatic driving,  which affects different tracers depending on their charge. Hence the corrugation, by modifying the velocity profile, acts as a selector that can enhance/reduce the convection of tracers depleted from the walls modulating the magnitude (and sign) of the tracer  flux  velocity. 
This geometrically-induced tuning, characteristic of the entropic electrokinetic regime,  can be quantified in terms of the competition between the characteristic  tracer electrophoretic velocity due to the applied electric field, $u_e= Z e \mu_{tr} \Delta{V}_0/L$, involving the tracer mobility $\mu_{tr}=\beta D_{tr}$, and the typical electroosmotic velocity the applied field induces in the solvent, $u_o = Q/h_0$ that will tend to carry  the tracers along. The inset of  Fig.~\ref{fig:tracers}.A  shows the ratio $u_e/u_o$ as a function of the channel corrugation, $\Delta S$, and quantifies the impact that the entropic restrictions induced by the inhomogeneous channel section has in diminishing the   effective drag induced by the electroosmotic flow. For very high values of $\Delta S$ the convective driving is reduced by the geometrically-enhanced dissipation leading to a reduction of the net fluid flux, $Q$. Moreover,  the electrostatic driving is affected by the geometrically-induced electrostatic barrier 
that 
forms, due to curvature, at the channel bottlenecks. 

Tracers of  radius $R$, due to their  finite size, experience an effective, size-dependent   corrugation, $\Delta S'=\ln\frac{h_0-R+h_1}{h_0-R-h_1}$,  larger than  the one affecting the electrolyte ($\Delta S' > \Delta S$). The restricted section sampled by the tracers, together with  the velocity  inversion in the channel center for electroosmotic flows, as shown in Fig.~\ref{fig:1}, enhances  mixing. Therefore, this  additional geometric constraint  leads to a complementary   mechanism of geometric, or entropic origin,  to control tracer  segregation    or tracer mixing at low Reynolds numbers according to tracer size and/or charge, as shown in Fig.~\ref{fig:tracers}.A. 

Fig.~\ref{fig:tracers}.B shows that for pressure driven flows  channel corrugation also decreases the tracer velocity, but does not lead to tracer inversion. Tracers attracted to the channel walls  are more reactive to channel corrugations and experience a faster decrease of their velocity increasing $\Delta S$. The asymmetry in the sensitivity of positive and negative tracers to channel spatial inhomogeneities leads to an inversion in the tracer electric current, as shown in the inset  of of Fig.~\ref{fig:tracers}.B.  This inversion emerges because the electrostatic potential is more susceptible to corrugation in the center of the channel, hence having a stronger impact on tracers  repelled by the channel walls.
This  is  an experimentally detectable phenomenon because the signal to noise ratio, captured by the tracer P\'eclet number $Pe_{tr}=u_0 h_0/D_{tr}$, can be of the order $\sim 1/10$ for $\Delta S\sim 3$. For example, this confinement-controlled current inversion will   efficiently separate nanometer-size particles with charge $\pm e$ on length scales of the order of $\sim 10 \mu m$ with micron size average amplitude channels and Debye length $\kappa^{-1}\sim 5 nm$. 

Local velocity inversion allows for negative mobility, where tracers displace on average against the net driving force~\cite{Eichhorn2010,Speer2012}. This phenomenon is more clearly appreciated  for neutral tracers, for which we can gain more insight because  local electric forces do not compete with  flow-induced drag. Tracers localized to the channel central section due the direct action of an external force (e.g. through optical trapping), will move against the mean electrolyte velocity  when  local flow inversion develops.  Since the regions of flow inversion are compact and do not  extend into the bottleneck regions, mobility inversion requires  tracers to diffuse over the narrow region channel  against the local flow to jump to the previous confined environment. As a result,  the mobility rapidly decreases   with the size  of the   inverse fluid flow region, as shown in  Fig.~\ref{fig:tracers}.C, and can invert the sign for larger corrugations.

The variable channel corrugation modifies  the tracer  transverse probability distribution along the channel and  enhances the role played by the central counter-flux leading to velocity inversion. Hence, corrugation induces a new kind of absolute negative mobility that does not require disorder, characterized by two different regimes. In the diffusive regime, $Pe_{tr}\ll 1$, the time required to displace along the central part of the channel due to the reversed flux, $\tau_{cnv}$, is the rate-limiting process, whereas in the convective regime, $Pe_{tr} \gtrsim 1$, the diffusion against the low in the channel bottlenecks, $\tau_{diff}$, becomes the limiting process.  Therefore, modifying the shape of the channel offers a means to  control the magnitude of the tracer negative mobility. For example,  tuning the size, $\delta$, of the central region of the channel it is possible to enlarge the relative relevance between $\tau_{diff}$ and $\tau_{conv}$ and therefore enhance the negative mobility, as shown in fig.
~\ref{fig:tracers}.C.

We have developed a flexible, theoretical  framework that has allowed us  to capture the essential dynamic properties of entropically-induced  electrokinetic  flow in a corrugated channel when the Debye length is comparable to the channel section. We have found that   electrokinetics in a spatially-varying channel section allows for a very versatile and sensitive regulation of the fluxes of both electrolytes and  tracers. We have clarified the relevance of  confinement for a $z-z$ electrolyte driven by a pressure gradient or an electric potential drop and have shown that the induced electrokinetic flows lead to a variety of new phenomena, such as  tracer  separation for electroosmosis, current inversion for pressure driven fluxes, and negative mobility for optically trapped neutral tracers in an electroosmotic flow. 
Entropically controlled electrokinetics can be achieved experimentally  in a variety of conditions. For example, ion transport  across nanometric porous media or the diffusion of radioactive ions in containers  takes place along nanometric-size matrix~\cite{Bertron201451}, where the Debye length is naturally of the same size. Dilute  electrolytes  can develop micrometric size Debye lengths, opening the possibility to study entropic  electrokinetics of macromolecules, proteins or aggregates in microfluidic devices. Finally, since channels with controllable corrugation have just been realized~\cite{def-channel}, entropic electrokinetics can provide an alternative way to the develop  nano- micro-fluidic devices.

We acknowledge  the Direcci\'on General de Investigaci\'on (Spain) and DURSI  for financial support
under projects  FIS\ 2011-22603 and 2009SGR-634, respectively. J.M.R. and I.P.  ackowledge financial support from {\sl Generalitat de Catalunya }under program {\sl Icrea Acad\`emia}.

\bibliography{letter_electrokinetics}

\begin{thebibliography}{30}
\expandafter\ifx\csname natexlab\endcsname\relax\def\natexlab#1{#1}\fi
\expandafter\ifx\csname bibnamefont\endcsname\relax
  \def\bibnamefont#1{#1}\fi
\expandafter\ifx\csname bibfnamefont\endcsname\relax
  \def\bibfnamefont#1{#1}\fi
\expandafter\ifx\csname citenamefont\endcsname\relax
  \def\citenamefont#1{#1}\fi
\expandafter\ifx\csname url\endcsname\relax
  \def\url#1{\texttt{#1}}\fi
\expandafter\ifx\csname urlprefix\endcsname\relax\def\urlprefix{URL }\fi
\providecommand{\bibinfo}[2]{#2}
\providecommand{\eprint}[2][]{\url{#2}}

\bibitem[{\citenamefont{Bocquet and Charlaix}(2010)}]{Lyderic}
\bibinfo{author}{\bibfnamefont{L.}~\bibnamefont{Bocquet}} \bibnamefont{and}
  \bibinfo{author}{\bibfnamefont{E.}~\bibnamefont{Charlaix}},
  \bibinfo{journal}{Chem. Soc. Rev.} \textbf{\bibinfo{volume}{39}},
  \bibinfo{pages}{1073} (\bibinfo{year}{2010}).

\bibitem[{\citenamefont{Alberts et~al.}(2007)\citenamefont{Alberts, Johnson,
  Lewis, Raff, Roberts, and Walter}}]{Albers}
\bibinfo{author}{\bibfnamefont{B.}~\bibnamefont{Alberts}},
  \bibinfo{author}{\bibfnamefont{A.}~\bibnamefont{Johnson}},
  \bibinfo{author}{\bibfnamefont{J.}~\bibnamefont{Lewis}},
  \bibinfo{author}{\bibfnamefont{M.}~\bibnamefont{Raff}},
  \bibinfo{author}{\bibfnamefont{K.}~\bibnamefont{Roberts}}, \bibnamefont{and}
  \bibinfo{author}{\bibfnamefont{P.}~\bibnamefont{Walter}},
  \emph{\bibinfo{title}{Molecular Biology of the Cell}}
  (\bibinfo{publisher}{Garland Science}, \bibinfo{address}{Oxford},
  \bibinfo{year}{2007}).

\bibitem[{\citenamefont{Karnik et~al.}(2007)\citenamefont{Karnik, Duan,
  Castelino, and Daiguji}}]{Karnik2007}
\bibinfo{author}{\bibfnamefont{R.}~\bibnamefont{Karnik}},
  \bibinfo{author}{\bibfnamefont{C.}~\bibnamefont{Duan}},
  \bibinfo{author}{\bibfnamefont{K.}~\bibnamefont{Castelino}},
  \bibnamefont{and} \bibinfo{author}{\bibfnamefont{H.}~\bibnamefont{Daiguji}},
  \bibinfo{journal}{Nano Letters} \textbf{\bibinfo{volume}{7}},
  \bibinfo{pages}{547} (\bibinfo{year}{2007}).

\bibitem[{\citenamefont{Park et~al.}(2006)\citenamefont{Park, Russo, Branton,
  and Stone}}]{Park2006}
\bibinfo{author}{\bibfnamefont{S.~Y.} \bibnamefont{Park}},
  \bibinfo{author}{\bibfnamefont{C.~J.} \bibnamefont{Russo}},
  \bibinfo{author}{\bibfnamefont{D.}~\bibnamefont{Branton}}, \bibnamefont{and}
  \bibinfo{author}{\bibfnamefont{H.~A.} \bibnamefont{Stone}},
  \bibinfo{journal}{Journal of Colloid and Interface Science}
  \textbf{\bibinfo{volume}{297}}, \bibinfo{pages}{832} (\bibinfo{year}{2006}).

\bibitem[{\citenamefont{Siria et~al.}(2013)\citenamefont{Siria, Poncharal,
  Biance1, Blase, Purcell, and Bocquet}}]{Lyderic_nature}
\bibinfo{author}{\bibfnamefont{A.}~\bibnamefont{Siria}},
  \bibinfo{author}{\bibfnamefont{P.}~\bibnamefont{Poncharal}},
  \bibinfo{author}{\bibfnamefont{R.}~\bibnamefont{Biance1},
  \bibfnamefont{A-L~Fulcrand}},
  \bibinfo{author}{\bibfnamefont{X.}~\bibnamefont{Blase}},
  \bibinfo{author}{\bibfnamefont{S.~T.} \bibnamefont{Purcell}},
  \bibnamefont{and} \bibinfo{author}{\bibfnamefont{L.}~\bibnamefont{Bocquet}},
  \bibinfo{journal}{Nature} \textbf{\bibinfo{volume}{494}},
  \bibinfo{pages}{455} (\bibinfo{year}{2013}).

\bibitem[{\citenamefont{Kosinska et~al.}(2008)\citenamefont{Kosinska, Goychuk,
  Kostur, Schmidt, and H\"{a}nggi}}]{hanggi}
\bibinfo{author}{\bibfnamefont{I.}~\bibnamefont{Kosinska}},
  \bibinfo{author}{\bibfnamefont{I.}~\bibnamefont{Goychuk}},
  \bibinfo{author}{\bibfnamefont{M.}~\bibnamefont{Kostur}},
  \bibinfo{author}{\bibfnamefont{G.}~\bibnamefont{Schmidt}}, \bibnamefont{and}
  \bibinfo{author}{\bibfnamefont{P.}~\bibnamefont{H\"{a}nggi}},
  \bibinfo{journal}{Phys. Rev. E} \textbf{\bibinfo{volume}{77}},
  \bibinfo{pages}{031131} (\bibinfo{year}{2008}).

\bibitem[{\citenamefont{Ghosal}(2002)}]{ghosal}
\bibinfo{author}{\bibfnamefont{S.}~\bibnamefont{Ghosal}}, \bibinfo{journal}{J.
  Fluid. Mech} \textbf{\bibinfo{volume}{459}}, \bibinfo{pages}{103}
  (\bibinfo{year}{2002}).

\bibitem[{\citenamefont{Dorfman}(2008)}]{Dorfman2008}
\bibinfo{author}{\bibfnamefont{K.}~\bibnamefont{Dorfman}},
  \bibinfo{journal}{Phys. Fluids} \textbf{\bibinfo{volume}{20}},
  \bibinfo{pages}{037102} (\bibinfo{year}{2008}).

\bibitem[{\citenamefont{Malevich et~al.}(2010)\citenamefont{Malevich,
  Mityushev, and Adler}}]{Adler}
\bibinfo{author}{\bibfnamefont{A.~E.} \bibnamefont{Malevich}},
  \bibinfo{author}{\bibfnamefont{V.}~\bibnamefont{Mityushev}},
  \bibnamefont{and} \bibinfo{author}{\bibfnamefont{P.~M.} \bibnamefont{Adler}},
  \bibinfo{journal}{J. Colloid Interface Sci.} \textbf{\bibinfo{volume}{345}},
  \bibinfo{pages}{72} (\bibinfo{year}{2010}).

\bibitem[{\citenamefont{Reguera et~al.}(2006)\citenamefont{Reguera, Schmid,
  Burada, Rub\'{\i}, Reimann, and H\"anggi}}]{Reguera2006}
\bibinfo{author}{\bibfnamefont{D.}~\bibnamefont{Reguera}},
  \bibinfo{author}{\bibfnamefont{G.}~\bibnamefont{Schmid}},
  \bibinfo{author}{\bibfnamefont{P.~S.} \bibnamefont{Burada}},
  \bibinfo{author}{\bibfnamefont{J.~M.} \bibnamefont{Rub\'{\i}}},
  \bibinfo{author}{\bibfnamefont{P.}~\bibnamefont{Reimann}}, \bibnamefont{and}
  \bibinfo{author}{\bibfnamefont{P.}~\bibnamefont{H\"anggi}},
  \bibinfo{journal}{Phys. Rev. Lett.} \textbf{\bibinfo{volume}{96}},
  \bibinfo{pages}{130603} (\bibinfo{year}{2006}).

\bibitem[{\citenamefont{Marini Bettolo~Marconi
  et~al.}(2013)\citenamefont{Marini Bettolo~Marconi, Melchionna, and
  Pagonabarraga}}]{Umberto_jcp}
\bibinfo{author}{\bibfnamefont{U.}~\bibnamefont{Marini Bettolo~Marconi}},
  \bibinfo{author}{\bibfnamefont{S.}~\bibnamefont{Melchionna}},
  \bibnamefont{and}
  \bibinfo{author}{\bibfnamefont{I.}~\bibnamefont{Pagonabarraga}},
  \bibinfo{journal}{J. Chem. Phys} \textbf{\bibinfo{volume}{138}},
  \bibinfo{pages}{244107} (\bibinfo{year}{2013}).

\bibitem[{\citenamefont{Martens et~al.}(2013)\citenamefont{Martens, Straube,
  Schmidt, Schimansky-Geier, and H\"{a}nggi}}]{Hanggi_PRL}
\bibinfo{author}{\bibfnamefont{S.}~\bibnamefont{Martens}},
  \bibinfo{author}{\bibfnamefont{A.~V.} \bibnamefont{Straube}},
  \bibinfo{author}{\bibfnamefont{G.}~\bibnamefont{Schmidt}},
  \bibinfo{author}{\bibfnamefont{L.}~\bibnamefont{Schimansky-Geier}},
  \bibnamefont{and}
  \bibinfo{author}{\bibfnamefont{P.}~\bibnamefont{H\"{a}nggi}},
  \bibinfo{journal}{Phys. Rev. Lett.} \textbf{\bibinfo{volume}{110}},
  \bibinfo{pages}{010601} (\bibinfo{year}{2013}).

\bibitem[{\citenamefont{Pagonabarraga et~al.}(2010)\citenamefont{Pagonabarraga,
  Rotenberg, and Frenkel}}]{Rotenberg_review}
\bibinfo{author}{\bibfnamefont{I.}~\bibnamefont{Pagonabarraga}},
  \bibinfo{author}{\bibfnamefont{B.}~\bibnamefont{Rotenberg}},
  \bibnamefont{and} \bibinfo{author}{\bibfnamefont{D.}~\bibnamefont{Frenkel}},
  \bibinfo{journal}{Phys. Chem. Chem. Phys} \textbf{\bibinfo{volume}{12}},
  \bibinfo{pages}{9566} (\bibinfo{year}{2010}).

\bibitem[{\citenamefont{Lemaire et~al.}(2005)\citenamefont{Lemaire, Naili, and
  Remond}}]{Lemaire2005}
\bibinfo{author}{\bibfnamefont{T.}~\bibnamefont{Lemaire}},
  \bibinfo{author}{\bibfnamefont{S.}~\bibnamefont{Naili}}, \bibnamefont{and}
  \bibinfo{author}{\bibfnamefont{A.}~\bibnamefont{Remond}},
  \bibinfo{journal}{Biomechan Model Mechanobiol} \textbf{\bibinfo{volume}{5}},
  \bibinfo{pages}{39} (\bibinfo{year}{2005}).

\bibitem[{\citenamefont{Wheeler and Stroock}(2008)}]{Strook2008}
\bibinfo{author}{\bibfnamefont{T.}~\bibnamefont{Wheeler}} \bibnamefont{and}
  \bibinfo{author}{\bibfnamefont{A.}~\bibnamefont{Stroock}},
  \bibinfo{journal}{Nature} \textbf{\bibinfo{volume}{455}},
  \bibinfo{pages}{208} (\bibinfo{year}{2008}).

\bibitem[{\citenamefont{Stroock et~al.}(2010)\citenamefont{Stroock, Dertinger,
  Ajdari, Mezic, Stone, and Whitesides}}]{Stroock2010}
\bibinfo{author}{\bibfnamefont{A.~D.} \bibnamefont{Stroock}},
  \bibinfo{author}{\bibfnamefont{S.~K.~W.} \bibnamefont{Dertinger}},
  \bibinfo{author}{\bibfnamefont{A.}~\bibnamefont{Ajdari}},
  \bibinfo{author}{\bibfnamefont{I.}~\bibnamefont{Mezic}},
  \bibinfo{author}{\bibfnamefont{H.~A.} \bibnamefont{Stone}}, \bibnamefont{and}
  \bibinfo{author}{\bibfnamefont{G.~M.} \bibnamefont{Whitesides}},
  \bibinfo{journal}{Science} \textbf{\bibinfo{volume}{647}}
  (\bibinfo{year}{2010}).

\bibitem[{\citenamefont{Yang et~al.}(2003)\citenamefont{Yang, Lu, Kostiuk, and
  Kwok}}]{Yang2003}
\bibinfo{author}{\bibfnamefont{J.}~\bibnamefont{Yang}},
  \bibinfo{author}{\bibfnamefont{F.}~\bibnamefont{Lu}},
  \bibinfo{author}{\bibfnamefont{L.~W.} \bibnamefont{Kostiuk}},
  \bibnamefont{and} \bibinfo{author}{\bibfnamefont{D.~Y.} \bibnamefont{Kwok}},
  \textbf{\bibinfo{volume}{963}} (\bibinfo{year}{2003}).

\bibitem[{\citenamefont{Picallo et~al.}(2013)\citenamefont{Picallo, Gravelle,
  Joly, Charlaix, and Bocquet}}]{Lyderic_PRL}
\bibinfo{author}{\bibfnamefont{C.~B.} \bibnamefont{Picallo}},
  \bibinfo{author}{\bibfnamefont{S.}~\bibnamefont{Gravelle}},
  \bibinfo{author}{\bibfnamefont{L.}~\bibnamefont{Joly}},
  \bibinfo{author}{\bibfnamefont{E.}~\bibnamefont{Charlaix}}, \bibnamefont{and}
  \bibinfo{author}{\bibfnamefont{L.}~\bibnamefont{Bocquet}},
  \bibinfo{journal}{Phys. Rev. Lett.} \textbf{\bibinfo{volume}{111}},
  \bibinfo{pages}{244501} (\bibinfo{year}{2013}).

\bibitem[{\citenamefont{Brogioli}(2009)}]{Brogioli2009}
\bibinfo{author}{\bibfnamefont{D.}~\bibnamefont{Brogioli}},
  \bibinfo{journal}{Phys. Rev. Lett.} \textbf{\bibinfo{volume}{103}},
  \bibinfo{pages}{058501} (\bibinfo{year}{2009}).

\bibitem[{\citenamefont{Taylor et~al.}(2011)\citenamefont{Taylor, Boon, and
  Roij}}]{Roji2011}
\bibinfo{author}{\bibfnamefont{P.}~\bibnamefont{Taylor}},
  \bibinfo{author}{\bibfnamefont{N.}~\bibnamefont{Boon}}, \bibnamefont{and}
  \bibinfo{author}{\bibfnamefont{R.~V.} \bibnamefont{Roij}},
  \bibinfo{journal}{Molecular Physics} \textbf{\bibinfo{volume}{109}},
  \bibinfo{pages}{37} (\bibinfo{year}{2011}).

\bibitem[{\citenamefont{Malgaretti et~al.}(2013)\citenamefont{Malgaretti,
  Pagonabarraga, and Rubi}}]{paolo_jcp_2013}
\bibinfo{author}{\bibfnamefont{P.}~\bibnamefont{Malgaretti}},
  \bibinfo{author}{\bibfnamefont{I.}~\bibnamefont{Pagonabarraga}},
  \bibnamefont{and} \bibinfo{author}{\bibfnamefont{J.~M.} \bibnamefont{Rubi}},
  \bibinfo{journal}{J. Chem. Phys.} \textbf{\bibinfo{volume}{138}},
  \bibinfo{pages}{194906} (\bibinfo{year}{2013}).

\bibitem[{\citenamefont{Zwanzig}(1992)}]{zwanzig}
\bibinfo{author}{\bibfnamefont{R.}~\bibnamefont{Zwanzig}}, \bibinfo{journal}{J.
  Phys. Chem.} \textbf{\bibinfo{volume}{96}}, \bibinfo{pages}{3926}
  (\bibinfo{year}{1992}).

\bibitem[{\citenamefont{Reguera and Rubi}(2001)}]{Reguera2001}
\bibinfo{author}{\bibfnamefont{D.}~\bibnamefont{Reguera}} \bibnamefont{and}
  \bibinfo{author}{\bibfnamefont{J.~M.} \bibnamefont{Rubi}},
  \bibinfo{journal}{Phys. Rev. E} \textbf{\bibinfo{volume}{64}},
  \bibinfo{pages}{061106} (\bibinfo{year}{2001}).

\bibitem[{\citenamefont{Kalinay and Percus}(2008)}]{Kalinay2008}
\bibinfo{author}{\bibfnamefont{P.}~\bibnamefont{Kalinay}} \bibnamefont{and}
  \bibinfo{author}{\bibfnamefont{J.~K.} \bibnamefont{Percus}},
  \bibinfo{journal}{Phys. Rev. E} \textbf{\bibinfo{volume}{78}},
  \bibinfo{pages}{021103} (\bibinfo{year}{2008}).

\bibitem[{\citenamefont{Rotenberg et~al.}(2008)\citenamefont{Rotenberg,
  Pagonabarraga, and Frenkel}}]{Rotenberg_epl}
\bibinfo{author}{\bibfnamefont{B.}~\bibnamefont{Rotenberg}},
  \bibinfo{author}{\bibfnamefont{I.}~\bibnamefont{Pagonabarraga}},
  \bibnamefont{and} \bibinfo{author}{\bibfnamefont{D.}~\bibnamefont{Frenkel}},
  \bibinfo{journal}{EPL} \textbf{\bibinfo{volume}{83}}, \bibinfo{pages}{34004}
  (\bibinfo{year}{2008}).

\bibitem[{\citenamefont{Eichhorn et~al.}(2010)\citenamefont{Eichhorn,
  Regtmeier, and Reimann}}]{Eichhorn2010}
\bibinfo{author}{\bibfnamefont{R.}~\bibnamefont{Eichhorn}},
  \bibinfo{author}{\bibfnamefont{J.}~\bibnamefont{Regtmeier}},
  \bibnamefont{and} \bibinfo{author}{\bibfnamefont{P.}~\bibnamefont{Reimann}},
  \bibinfo{journal}{Soft Matter} \textbf{\bibinfo{volume}{6}},
  \bibinfo{pages}{1858} (\bibinfo{year}{2010}).

\bibitem[{\citenamefont{Speer et~al.}(2012)\citenamefont{Speer, Eichhorn,
  Evstigneev, and Reimann}}]{Speer2012}
\bibinfo{author}{\bibfnamefont{D.}~\bibnamefont{Speer}},
  \bibinfo{author}{\bibfnamefont{R.}~\bibnamefont{Eichhorn}},
  \bibinfo{author}{\bibfnamefont{M.}~\bibnamefont{Evstigneev}},
  \bibnamefont{and} \bibinfo{author}{\bibfnamefont{P.}~\bibnamefont{Reimann}},
  \bibinfo{journal}{Phys. Rev. E} \textbf{\bibinfo{volume}{85}},
  \bibinfo{pages}{061132} (\bibinfo{year}{2012}).

\bibitem[{\citenamefont{Bertron et~al.}(2014)\citenamefont{Bertron, Jacquemet,
  Erable, Sablayrolles, Escadeillas, and Albrecht}}]{Bertron201451}
\bibinfo{author}{\bibfnamefont{A.}~\bibnamefont{Bertron}},
  \bibinfo{author}{\bibfnamefont{N.}~\bibnamefont{Jacquemet}},
  \bibinfo{author}{\bibfnamefont{B.}~\bibnamefont{Erable}},
  \bibinfo{author}{\bibfnamefont{C.}~\bibnamefont{Sablayrolles}},
  \bibinfo{author}{\bibfnamefont{G.}~\bibnamefont{Escadeillas}},
  \bibnamefont{and} \bibinfo{author}{\bibfnamefont{A.}~\bibnamefont{Albrecht}},
  \bibinfo{journal}{Nuclear Engineering and Design}
  \textbf{\bibinfo{volume}{268}}, \bibinfo{pages}{51} (\bibinfo{year}{2014}).

\bibitem[{\citenamefont{Chakraborty et~al.}(2012)\citenamefont{Chakraborty,
  Prakash, Friend, and Yeo}}]{def-channel}
\bibinfo{author}{\bibfnamefont{D.}~\bibnamefont{Chakraborty}},
  \bibinfo{author}{\bibfnamefont{J.~R.} \bibnamefont{Prakash}},
  \bibinfo{author}{\bibfnamefont{J.}~\bibnamefont{Friend}}, \bibnamefont{and}
  \bibinfo{author}{\bibfnamefont{L.}~\bibnamefont{Yeo}},
  \bibinfo{journal}{Phys. of Fluids} \textbf{\bibinfo{volume}{24}},
  \bibinfo{pages}{102002} (\bibinfo{year}{2012}).

\bibitem[{\citenamefont{Gillespie et~al.}(2002)\citenamefont{Gillespie, Nonner,
  and Eisenberg}}]{Eisenberg}
\bibinfo{author}{\bibfnamefont{D.}~\bibnamefont{Gillespie}},
  \bibinfo{author}{\bibfnamefont{W.}~\bibnamefont{Nonner}}, \bibnamefont{and}
  \bibinfo{author}{\bibfnamefont{R.~S.} \bibnamefont{Eisenberg}},
  \bibinfo{journal}{Journal of Physics: Condensed Matter}
  \textbf{\bibinfo{volume}{14}}, \bibinfo{pages}{12129} (\bibinfo{year}{2002}).

\end{thebibliography}

\end{document}